\documentclass[amsmath, amssymb, aps, pra, onecolumn,
groupedaddress, superscriptaddress, preprint]{revtex4-1}

\usepackage{graphicx}
\usepackage{xcolor}

\begin{document}
\title{Physical random bit generation from chaotic solitary laser diode}

\author{Martin Virte}%
\email{martin.virte@supelec.fr}
\affiliation{Sup\'{e}lec, OPTEL Research Group, Laboratoire Mat\'{e}riaux
Optiques, Photonique et Syst\`{e}mes (LMOPS) EA-4423, 2 Rue Edouard Belin,
F-57070 Metz, France}
\affiliation{Brussels Photonics Team, Dept. of Applied Physics and
Photonics (B-PHOT TONA), Vrije Universiteit Brussels, Pleinlaan 2,
1050 Brussels}

\author{Emeric Mercier}%
\affiliation{Sup\'{e}lec, OPTEL Research Group, Laboratoire Mat\'{e}riaux
Optiques, Photonique et Syst\`{e}mes (LMOPS) EA-4423, 2 Rue Edouard Belin,
F-57070 Metz, France}
\affiliation{Brussels Photonics Team, Dept. of Applied Physics and
Photonics (B-PHOT TONA), Vrije Universiteit Brussels, Pleinlaan 2,
1050 Brussels}

\author{Hugo Thienpont}
\affiliation{Brussels Photonics Team, Dept. of Applied Physics and
Photonics (B-PHOT TONA), Vrije Universiteit Brussels, Pleinlaan 2,
1050 Brussels}

\author{Krassimir Panajotov}
\affiliation{Brussels Photonics Team, Dept. of Applied Physics and
Photonics (B-PHOT TONA), Vrije Universiteit Brussels, Pleinlaan 2,
1050 Brussels}
\affiliation{Institute of Solid State Physics, 72 Tzarigradsko Chaussee
Blvd., 1784 Sofia, Bulgaria}

\author{Marc Sciamanna}
\affiliation{Sup\'{e}lec, OPTEL Research Group, Laboratoire Mat\'{e}riaux
Optiques, Photonique et Syst\`{e}mes (LMOPS) EA-4423, 2 Rue Edouard Belin,
F-57070 Metz, France}


\begin{abstract}
We demonstrate the physical generation of random bits at  high bit rates ($>$ 100 Gb/s) using optical chaos from a solitary laser diode and therefore without the complex addition of either external optical feedback or injection. This striking result is obtained despite the low dimension and relatively small bandwidth of the laser chaos, i.e. two characteristics that have been so far considered as limiting the performances of optical chaos-based applications. We unambiguously attribute the successful randomness at high speed to the physics of the laser chaotic polarization dynamics and the resulting growth rate of the dynamical entropy.
\end{abstract}


\maketitle

\section{Introduction}
Due to increasing needs for secure communication and cryptography, various concepts of physical Random Number Generators (RNGs) have been suggested, such as entangled systems \cite{Pironio2010}, quantum fluctuations \cite{Gabriel2010, Symul2011}, weakly coupled super-lattices \cite{Li2013a}, or amplified spontaneous emission \cite{Liu2013, Williams2010}. Of these concepts, the RNGs based on optical chaos have so far shown the most promising results especially in terms of speed with bit rates up to tens [7--10] and even hundreds of Gb/s [11--15]. Current implementations are based on external time-delayed feedback and/or optical injection to generate chaos from a laser diode. Their performances are intimately linked to the following requirements: a broad and flat power spectrum [7, 15--17], a fast autocorrelation decay \cite{Oliver2011, Oliver2013}, a fast increase of the entropy \cite{Mikami2012}, and positive finite-time Lyapunov exponents \cite{Reidler2009}. Consequently, these requirements motivate the use of rather strong  optical feedback strength \cite{Oliver2011}, hence of complex high-dimensional chaos \cite{Fischer1994, Vicente2005}. \\

Recently we have reported on chaos from a laser diode without external perturbation or forcing \cite{Virte2012}. Such a solitary single micro cavity laser diode, driven by a DC current, generates chaos at high speed and has the potential to greatly simplify optical systems used for chaos-based applications. However the resulting low chaos dimension and small number of positive Lyapunov exponents may cast doubts on its applicability for e.g. random number generation or cryptography where randomness at high speed is required.\\
Although many recent studies have reported tremendous performances, the main focus was on increasing the dynamics bandwidth \cite{Oliver2011, Yamazaki2013}, finding the proper optimization process \cite{Oliver2013} or the best post-processing method \cite{Reidler2009, Kanter2009, Li2013,  Yamazaki2013}. By contrast, how the characteristics of the chaotic dynamics - dimension, entropy evolution, Lyapunov exponent spectrum - impacts on the performances remains scarcely studied \cite{Mikami2012}. In this contribution, we use adistinctly different chaotic dynamics originating from polarization mode competition. The chaotic polarization fluctuations generated by the free-running laser diode show characteristics that differ significantly from those of the typical dynamics used so far for random bit generation (mostly chaotic dynamics generated by delayed optical feedback): polarization chaos is of low dimension, the polarization chaotic fluctuations show a single-peaked auto-correlation and therefore no correlation at higher times such as in time-delayed external-cavity laser diode, and finally the polarization chaos involves undamping of the laser diode relaxation oscillations (high frequencies) but with the spectral signature of the polarization mode-hopping at lower frequencies. Although the fast decaying single-peaked autocorrelation is as an advantage for random bit generation in contrast to delayed feedback laser diodes, the small chaos dimension and the limited bandwidth are commonly considered as detrimental for extracting randomness at high speed from the chaotic time-trace [7, 15--17]. Nevertheless, we successfully achieve close to the state-of-the-art performances ($>$ 100 Gbps) using this low-dimension and low-bandwidth dynamics without complex or unconventional processing. We therefore highlight that the performances of optical chaos-based RNG rely more on how fast is produced Shannon information theory entropy than on having a large chaos dimension and/or a large chaos bandwidth. This conclusion is supported by a theoretical analysis of the Shannon entropy growth rate from simulated polarization dynamics. We show theoretically that the low-pass filtering at 2.4 GHz as implemented experimentally actually plays a crucial role in producing fast entropy growth, and we link this finding to the filtering of the fast dynamical transitions in the double-scroll chaotic attractor that characterizes polarization chaos.\\
Our main conclusions are therefore that i) a low dimensional chaos can be efficiently used for RNGs, ii) competitive performance, up to 100 Gb/s with a minimal post-processing, can therefore be achieved using polarization chaos generated directly from a solitary laser diode, iii) such performance can be achieved with narrow-bandwidth equipment as opposed to the large bandwidth commonly required. We theoretically show that those three features directly result from the physics of the  double-scroll attractor of polarization chaos. The demonstration of competitive performances using a chaotic dynamics with at first sight not optimal properties and the unintuitive impact of a low-pass filtering in the context of random number generation at high-speed clearly motivates a new direction of research. Indeed, instead of improving chaos bandwidth or dynamical complexity,  we suggest to exploit how the specific features of the dynamics impact on the Shannon entropy growth rate to improve the performances of RNGs. In addition, the proposed scheme potentially opens the way toward heavy on chip-parallelization which might not be possible with other, more complex, schemes \cite{Argyris2010}.\\

\begin{figure*}[t]
		\includegraphics[width=\linewidth]{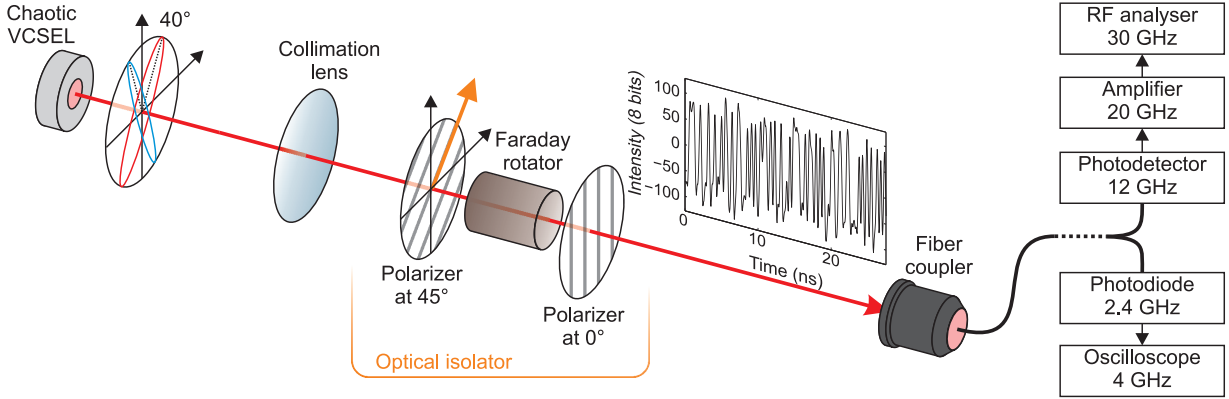}
		\caption{Experimental setup, from polarization chaos to chaotic time series. The VCSEL generates polarization chaos: using a polarizer at $45^{\circ}$ with respect to the linear polarization direction at threshold, polarization chaos can be transformed into chaotic intensity fluctuations. This polarizer combined with a rotator and a second polarizer form an optical isolator which prevents reflections from the fibre. RF spectra are acquired with a fast photo detector, an electronic amplifier and a RF spectrum analyser, whereas time-series are recorded with the small-bandwidth photodiode of the oscilloscope optical channel. Electronic bandwidths are indicated for each device. \label{fig:setup}}
\end{figure*}

\section{Experimental setup: random bit generation scheme}
The key element in our scheme is a vertical-cavity surface-emitting laser (VCSEL) generating polarization chaos \cite{Virte2012}. A detailed description of the device can be found in [22--24]. 
The device emits with a linearly polarized (LP) light at threshold but when increasing the current it bifurcates to emission along one of two elliptically polarized light states with their main axes being separated by about 40 degrees. A further increase of current destabilizes these elliptically polarized states steady-states successively to self-pulsating dynamics, quasiperiodicity and chaos \cite{Virte2012}. The linear birefringence of the VCSEL has been measured as about 5.6 GHz \cite{Olejniczak2011}. As demonstrated theoretically \cite{Virte2013} the linear birefringence is an important parameter explaining the bifurcation scenario observed experimentally. The chaotic dynamics consists of a random-like hopping between the two elliptically polarized states – represented by the red and blue ellipses in Fig. \ref{fig:setup}. To convert this polarization chaos into chaotic intensity fluctuations we use a polarizer oriented at $45^{\circ}$ with respect to the linear polarization direction at threshold so that mode hopping is achieved with the largest amplitude \cite{Olejniczak2011}. A typical time-trace of the chaotic fluctuations is given in Fig. \ref{fig:setup}. As in any other scheme generating and sampling optical chaos from a laser diode, we use an optical isolator to avoid any unintentional feedback from the measurement devices, in particular in our case undesired reflections from the fiber front-facet. The fiber is then plugged into two different measurement branches: one for time-series acquisition using the small-bandwidth photodiode (2.4 GHz bandwidth) of the oscilloscope optical channel (Tektronix CSA7404, 4GHz, 20 GS/s) and the other for RF spectrum acquisitions using a high-bandwidth photo detector (NewFocus 1554-B, 12 GHz), a RF amplifier (NewFocus 1422-LF, 20 GHz) and a RF spectrum analyzer (Anritsu MS2667c, 7kHz-30GHz). We furthermore use a variable neutral density filter to adjust the optical power received by the small-bandwidth photodiode in order to fully use the 8-bit range of the oscilloscope ADC - we typically target a ratio of saturated points between $10^{-4}$ and $10^{-1}$ \% estimated over 2 million points.\\
In the end, the only requirements of our optical chaos-based RNG is to add to the collimated laser output a polarizer to make a projection of the dynamics into the appropriate phase space and to adjust temperature and injection current to obtain polarization chaos. This chaos-based RNG is therefore kept at its utmost simplicity when comparing to more complex configurations relying on external feedback and/or optical injection.\\

\begin{figure}[t]
	\begin{center}
		\includegraphics[width=0.6\linewidth]{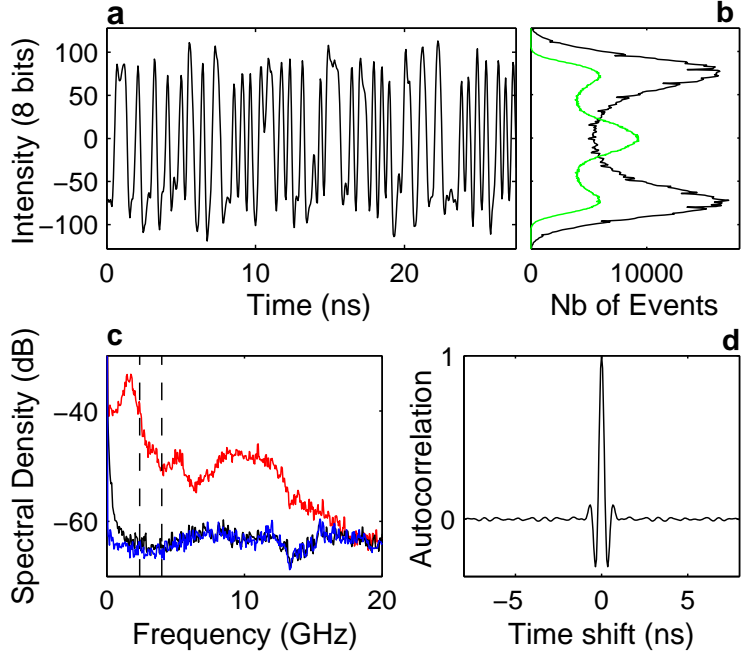}
		\caption{Characterization of Polarization Chaos Time-Series. (a) Typical intensity fluctuations recorded by the oscilloscope 8-bit ADC and the small-bandwidth photodiode with the corresponding distribution given by the black curve in (b); the green curve shows the resulting histogram after a comparison of the time-series with a time-shifted (250 ps) version of itself. (c) RF spectra of the chaotic fluctuations, laser noise and amplification noise in red, black and blue respectively. Vertical dotted lines give the electronic bandwidth of the small-bandwidth photodiode of the oscilloscope optical channel and of the oscilloscope (2.4 and 4 GHz resp.). (d) Autocorrelation of the time-series. \label{fig:chaos}}
	\end{center}
\end{figure}

Polarization chaos yields a double-scroll butterfly attractor that shares many similar properties to Lorenz chaos, including a similar value of the correlation dimension and a similar scaling of the dwell time in the chaotic switching between the two scrolls as a function of the "r" variable (in Lorenz) or pump parameter (in VCSEL). See e.g. Figs. 2 and 3 in \cite{Virte2012}. As a result of the double-scroll attractor, we can identify two different time-scales in the dynamics: the rapid one being the oscillation frequency on each of the two attractor wings, the slow one being the time between two successive jumps from one wing to the other, also called the dwell-time. A typical time-trace of the slow mode hopping dynamics is shown in Fig. \ref{fig:chaos}(a) along with the corresponding two-peak histogram in Fig. \ref{fig:chaos}(b). We measure an average dwell-time of about 0.6 ns. Both slow and fast dynamical time-scales can be distinguished in the RF spectra in Fig. \ref{fig:chaos}(c): low-frequencies correspond to the mode hopping, with a relatively broad peak centered around 1.7 GHz, whereas the top-hat-shaped peak around 10 GHz is induced by the fast oscillations around the unstable elliptically polarized states \cite{Virte2012}. The autocorrelation of the recorded time-series is shown in Fig. \ref{fig:chaos}(d), featuring a single oscillation and a large central peak. The width of the central peak is about 0.7 ns, i.e. two times larger than for the chaotic dynamics reported in RNG systems based on time-delayed feedback \cite{Uchida2008, Oliver2011}. Because of the absence of time delay, no correlation appears however for time-shifts above 1 ns.  By applying the Grassberger-Procaccia algorithm, we obtain a correlation dimension of $C_D \approx 2.3$,  which is similar to the one reported in \cite{Virte2012}, and confirms the low dimension of the chaos.  It is worth emphasizing that the property of the dynamics, such as its deterministic nature or its dimension, is conserved when applying operations like a projection (as performed by the polarizer) or a filtering (as performed by the photodiode and oscilloscope).\\

\section{Analysis of polarization chaos dynamics for random bit generation}
\begin{figure}[htb]
	\begin{center}
		\includegraphics[width=0.6\linewidth]{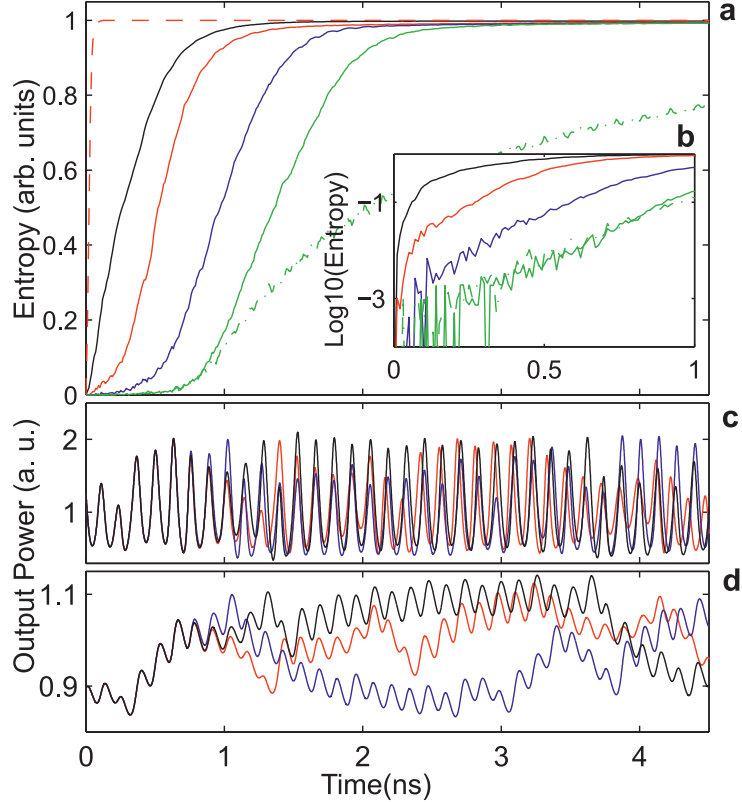}
		\caption{Entropy evolution for polarization chaos. (a) Black, red, blue and green give the entropy evolution for noise levels of -30, -40, -50 and -60 dB respectively. Dash-doted, solid and dashed lines give the evolution for unfiltered, filtered, and filtered + processed cases respectively. (b) is the same plot as (a) but in logarithmic scale for the entropy. (c) and (d) gives an example of the separation between 3 different chaotic time-series with a unfiltered (c) and filtered output power (d) and noise level of -60 dB. \label{fig:entropy}}
	\end{center}
\end{figure}

Our polarization dynamics showing sensitivity to initial conditions, any intrinsic noise as practically present in our device will be amplified by the dynamical instability such that the polarization state after some transient time cannot be determined. The polarization output can therefore be considered as not only unpredictable but also statistically random \cite{Fox1991, Bracikowski1992}. As suggested in \cite{Mikami2012}, it is crucial at this stage to estimate the rate at which the non-deterministic random bits can be generated from the noise amplification by the chaotic dynamics. Such an estimation is difficult to obtain from the experimental time-series since it means looking at the system dynamics from the same initial state. However following the procedure of \cite{Mikami2012}, we can analyze the evolution of the dynamical entropy from the simulation of a large number of polarization output time-series starting from the same initial conditions but with different noise sequences. As we reported elsewhere \cite{Virte2012, Virte2013}, the experimentally recorded polarization chaos is qualitatively well reproduced using the spin-flip rate equation model (SFM) \cite{SanMiguel1995, Martin-Regalado1997}. We use the same set of parameters as in \cite{Virte2012, Virte2013}. In particular, we take an injection current of $\mu = 2.963$ and a birefringence of $\gamma_p = 25$ ns\textsuperscript{-1} for which polarization chaos is observed with fast pulsations at a frequency of around 8 GHz and with an averaged polarization switching dwell time of 0.6 ns. The equipment limited bandwidth is modeled by a low-pass first-order filter with cut-off frequency at 2.4 GHz. The time-evolution of the dynamical entropy is shown in Fig. \ref{fig:entropy}. The comparison between the dash-dotted line for the un-filtered and the solid lines for the filtered polarization dynamics shows a significant effect of the limited detection bandwidth on the dynamical entropy. This is confirmed in Fig. \ref{fig:entropy}(c)--\ref{fig:entropy}(d) which plot the trajectories of the output power for three different noise sequences starting from the same initial condition, for the un-filtered and the filtered dynamics respectively. The trajectories in both cases diverge because of the amplification of the noise by the chaotic dynamics but filtering the fast dynamics of our chaos – i.e. the rapid and strongly dominating oscillations – greatly helps to observe a macroscopic divergence more quickly. As suggested in \cite{Mikami2012}, we define the memory time as the time when the entropy reaches the value of 0.995, i.e. the time after wich it can be considered that the information about the initial state of the system is lost. Figure \ref{fig:entropy}(a) shows that the filtered chaotic dynamics exhibits more than eight times shorter memory times for similar noise levels than the one reported in the time-delayed feedback case \cite{Mikami2012} : around 1.5 ns in Fig. \ref{fig:entropy} versus more than 13 ns respectively at a -30 dB noise level. This short memory time is also explained by the large value, around 7 ns\textsuperscript{-1}, of the largest Lyapunov exponent theoretically determined with Shimada's algorithm \cite{Shimada1979}. The effect of the Lyapunov Exponent is visible in Fig. \ref{fig:entropy}(b) as it corresponds to the slope of the entropy increase versus time \cite{Mikami2012}. We also find an entropy rate - which evaluates the memory time evolution with respect to the noise level - of 4.8 ns\textsuperscript{-1} to be compared with the 1.7 ns\textsuperscript{-1} value obtained in \cite{Mikami2012}. Considering these theoretical results, we could expect to obtain random sequences using a single threshold level and a sampling time larger than 1.5 ns - i.e sampling rate up to 660 MS/s. It should however be clearly emphasized that, as shown above, such short memory times could not be achieved without the low-pass filtering dismissing the fast oscillations of the dynamics.\\

The previous theoretical analysis motivates the use of our polarization chaos as a physical source for a non-deterministic and rapidly varying entropy. But similarly to previous experimental realizations of chaos-based RNG [7, 9, 11--17], a suitable post-processing must be implemented to first overcome the weaknesses of the entropy source, in particular to remove the bias in the sequence of zeros and ones, and secondly to possibly increase the random bit rate. As already reported \cite{Uchida2008}, using one threshold to generate random sequences is quite challenging as it needs to be precisely set and readjusted to avoid long-term bias. To reduce experimental constraints, previous studies have suggested multibit generation schemes, generally combined with bit truncation \cite{Reidler2009, Yamazaki2013, Oliver2013}. Here, however, using bit truncation methods alone results in a strongly biased sequence even when considering only the least significant bit (typ. 50.1\% of 1 versus 49.9\% of 0 for sequences of 2 million bits, i.e. $P\_value \sim 4e-12$). In order to compensate for the statistical imperfection, we suggest to compare recorded time trace with a time-shifted version of itself. The time-shift is set at 250 ps which corresponds to the time for the decay of the autocorrelation. The time-shift is therefore large enough to avoid correlated dynamics but is also small enough to keep the details of the chaotic dynamics and mostly the chaotic hopping between polarization states that occur with an averaged dwell time of 0.5 ns. The histogram of the polarized output power after the comparison (green curve in Fig. \ref{fig:chaos}(b)) shows a more symmetric distribution for producing an unbiased sequence of random bits. Additionally, as shown by the dashed curve of Fig. \ref{fig:entropy}(a), implementing a single comparison on a multibit extraction scheme allows to significantly decrease the system memory time down to about 0.05 ns, hence suggesting a sampling rate up to about 20 GS/s for the time-series acquisition.\\

\section{Random bit extraction: processing and experimental results}
To evaluate the randomness of the final bit sequence we use two standard batteries of tests: the NIST statistical test suite \cite{Rukhin2010}, and the Diehard test suite \cite{Marsaglia1996}. Using the post-processing described previously, we successfully generate sequences that pass all the NIST and DieHard test by retaining up to 5 LSBs for a waveform recorded at 20 GS/s and with a time-shift of 250 ps; the test results are given in Table \ref{tab:NIST} and \ref{tab:dieHard}, case 1. We have therefore experimentally demonstrated that the low-dimensional polarization chaos can be efficiently used to produce random sequences at a highly competitive bit rate (5 bits x 20 GS/s = 100 Gb/s) without unconventional and/or complex post-processing methods. These performances are in good agreement with the theoretical expectations based on the SFM model. Additionally, we have checked that a much higher bit rate can be achieved by increasing the number of comparisons between the time-shifted waveforms, hence allowing keeping more bits for each data point \cite{Reidler2009}. When increasing the number of retained bits up to 28 when performing 23 comparisons with the same time-shift of 250 ps, we generate a bit sequence, that passes all the NIST tests as shown in Table \ref{tab:NIST} and \ref{tab:dieHard} case 2, with a potential bitrate of 560 Gb/s.\\

\section{Conclusion}
To conclude, we report on the physical generation of random numbers at high-speed from the sampling of the polarization chaotic dynamics of a solitary laser diode, or in other words a fast optical chaos-based random number generator (RNG) that does not require any external optical feedback or coupling. The polarization chaotic dynamics shows the properties of a low-dimension double scroll chaotic attractor, with chaotic switchings between two unstable fixed points that correspond to elliptically polarized light states. We prove that a fast rising entropy and a therefore short memory times (about 1.5 ns in the case considered) can surprisingly be achieved by low-pass filtering the polarization dynamics with a cut-off frequency slightly larger than the frequency of the polarization switchings (measured as 1.7 GHz). Without the low-pass filter we observe memory times about 10 times larger, hence confirming its crucial impact on the entropy growth. This physical property of the chaotic dynamics makes it possible to extract random bits at above Gbit/s in spite of the therefore low-dimension of the chaotic attractor and in spite of the relatively small bandwidth. We demonstrate 100 Gb/s random bit generation passing successfully all NIST and Diehard statistical tests using a minimal post-processing (a single comparison and retaining the 5 least significant bits), and even 560 Gb/s with the use of high-order derivatives. Our scheme for an optical chaos-based RNG makes use of a new physical source of entropy, i.e. chaotic polarization switching, and in spite of its simplicity (a DC-current driven single laser diode) shows comparable or even higher performances than existing schemes based on chaotic laser dynamics from either optical feedback or external optical injection. In addition, we theoretically show that as opposed to the commonly accepted requirement - that links larger bandwidth to faster RBGs - we could benefit from a low-pass filtering of the dynamics; this interpretation is strongly supported by experimental results in agreement with the theoretical expectations. Finally, our scheme also benefits from the advantages of using a solitary VCSEL as an optical system, in particular low-cost, low consumption (mA driving current) and potential parallelization based on two-dimensional VCSEL arrays.\\

\section*{Acknowledgments}
The authors acknowledge support from the Conseil R\'{e}gional de Lorraine, Fondation Sup\'{e}lec, FWO-Vlaanderen, the METHUSALEM programme of the Flemish government, and the interuniversity attraction poles programme of the Belgian Science Policy Office (grant no. IAP P7-35 ``Photonics@be'').

\begin{table}[p]
\center
\caption{Results of the NIST Statistical Test. The tests were performed with 1000 sequences of 1 million bits each. We use a significance level of $\alpha = 0.01$, hence the tests are successful if the P-value (uniformity of the p-values) is larger than $0.0001$ and the proportion is in the range $0.99 \pm 0.0094392$. For tests generating multiple outputs we give the worst case scenario. In case 1, we use a single comparison and keep the 5 LSBs of the waveform. In case 2, we perform 23 comparisons and keep 23 LSBs of the waveform. In both case the time-shift is 250 ps.}
\begin{tabular}{l | c c | c c |}
\cline{2-5}
  & \multicolumn{2}{c |}{Case 1 } & \multicolumn{2}{c |}{Case 2} \\
\bf Statistical Tests & \bf P-Value & \bf Prop. & \bf P-value & \bf Prop. \\
\hline \hline
Frequency & 0.0127 & 0.9861 & 0.0170 & 0.9881\\
Block-Frequency & 0.5687 & 0.9851 & 0.0160 & 0.9921 \\
Cumulative Sum & 0.0023 & 0.9831 & 0.0954 & 0.9881 \\
Runs & 0.7116 & 0.9821 & 0.7637 & 0.9891 \\
Longest Run & 0.4172 & 0.9871 & 0.2034 & 0.9891 \\
Rank & 0.9819 & 0.9881 & 0.0127 & 0.9921 \\
FFT & 0.8360 & 0.9901 & 0.2284 & 0.9812 \\
Non Overlap. Templates & 0.0061 & 0.9812 & 0.0029 & 0.9812 \\
Overlapping Templates & 0.1007 & 0.9901 & 0.8920 & 0.9901 \\
Universal & 0.2480 & 0.9851 & 0.0661 & 0.9911 \\
Approximate Entropy & 0.0037 & 0.9861 & 0.6475 & 0.9921 \\
Random Excursions & 0.0354 & 0.9832 & 0.2858 & 0.9873 \\
Random Excursions Var. & 0.2162 & 0.9866 & 0.0038 & 0.9825 \\
Serial & 0.0767 & 0.9861 & 0.3161 & 0.9931 \\
Linear Complexity & 0.6579 & 0.9901 & 0.3283 & 0.9891 \\
\hline \hline
\end{tabular}
\label{tab:NIST}
\end{table}

\begin{table}[p]
\center
\caption{Results of the Diehard Statistical Test. The tests were performed with a sequence of 80 millions bits. For tests with multiple P-value worst case was selected. The tests are considered successful if the P-value is between 0.01 and 0.99. Description of case 1 and 2 is similar to the one given for Table \ref{tab:NIST}. $[KS]$ means that a Kolmogorov-Sinai test was applied. }
\begin{tabular}{l | c c | c c |}
\cline{2-5}
  & \multicolumn{2}{c |}{Case 1 } & \multicolumn{2}{c |}{Case 2} \\
\bf Statistical Tests & \bf P-Value & \bf Success & \bf P-value & \bf Success \\
\hline \hline
Birthday Spacing [KS] & 0.956792 & success & 0.389664 & success\\
Overlapping 5-permutation & 0.186900 & success & 0.128625 & success\\
Binary Rank for 31x31 matrices & 0.872671 & success & 0.912627 & success\\
Binary Rank for 32x32 matrices & 0.327389 & success & 0.382529 & success\\
Binary Rank for 6x8 matrices [KS] & 0.618094 & success & 0.538768 & success\\
Bistream & 0.08572 & success & 0.11933 & success\\
Overlapping-Pairs-Sparse-Occupancy & 0.0112 & success & 0.0468 & success\\
Overlapping-Quadruple-Sparse & 0.0462 & success & 0.0254 & success\\
DNA & 0.0831 & success & 0.0854 & success\\
Count the 1s on a stream of bytes & 0.360770 & success & 0.722437 & success\\
Count the 1s for specific bytes & 0.040344 & success & 0.022000 & success\\
Parking lot [KS] & 0.850322 & success & 0.784046 & success\\
Minimum Distance [KS] & 0.266732 & success & 0.179058 & success\\
3D spheres [KS] & 0.175171 & success & 0.690888 & success\\
Squeeze & 0.825330 & success & 0.332211 & success\\
Overlapping Sums [KS] & 0.161946 & success & 0.036895 & success\\
Runs [KS] & 0.172702 & success & 0.134305 & success\\
Craps & 0.303552 & success & 0.429254 & success\\
\hline \hline
\end{tabular}
\label{tab:dieHard}
\end{table}

%

\end{document}